\documentclass[pra,aps,showpacs,showkeys,superscriptaddress,11pt]{revtex4}
\usepackage{amsmath,amsfonts,theorem,epsfig}

\newcommand{\ket}[1]{| #1 \rangle}
\newcommand{\bra}[1]{\langle #1 |}

\newcommand{\proj}[1]{| #1\rangle\!\langle #1 |}
\newcommand{\Tr}{{\rm Tr}}

\def\qed{\ifmmode$\Box$\else{\unskip\nobreak\hfil
\penalty50\hskip1em\null\nobreak\hfil$\Box$
\parfillskip=0pt\finalhyphendemerits=0\endgraf}\fi}
\def\endenv{\ifmmode\;\else{\unskip\nobreak\hfil
\penalty50\hskip1em\null\nobreak\hfil\;
\parfillskip=0pt\finalhyphendemerits=0\endgraf}\fi}

\newcommand{\nc}{\newcommand}

\nc{\cA}{{\cal A}}
\nc{\cB}{{\cal B}}
\nc{\cC}{{\cal C}}
\nc{\cD}{{\cal D}}
\nc{\cE}{{\cal E}}
\nc{\cF}{{\cal F}}
\nc{\cG}{{\cal G}}
\nc{\cH}{{\cal H}}
\nc{\cI}{{\cal I}}
\nc{\cJ}{{\cal J}}
\nc{\cK}{{\cal K}}
\nc{\cL}{{\cal L}}
\nc{\cQ}{{\cal Q}}
\nc{\cR}{{\cal R}}
\nc{\cS}{{\cal S}}
\nc{\cX}{{\cal X}}

\nc{\dg}{\dagger}
\nc{\supp}{{\operatorname{supp}}}
\nc{\rar}{\rightarrow}
\nc{\lrar}{\longrightarrow}
\nc{\smfrac}[2]{\mbox{$\frac{#1}{#2}$}}
\nc{\ox}{\otimes}
\nc{\id}{\mathbb{I}}

\def\a{\alpha}

\def\d{\delta}
\def\e{\epsilon}

\def\o{\omega}

\def\r{\rho}
\def\s{\sigma}

\def\ph{\varphi}

\def\D{\Delta}

\def\Ph{\Phi}

\nc{\RR}{{{\mathbb R}}}
\nc{\CC}{{{\mathbb C}}}
\nc{\FF}{{{\mathbb F}}}
\nc{\NN}{{{\mathbb N}}}
\nc{\ZZ}{{{\mathbb Z}}}
\nc{\PP}{{{\mathbb P}}}
\nc{\QQ}{{{\mathbb Q}}}
\nc{\UU}{{{\mathbb U}}}
\nc{\Q}{{{\bar{Q}}}}

\begin{document}

\title{{\Large Hiding Quantum Data}}
\author{David P. DiVincenzo}
\email{divince@watson.ibm.com}
\affiliation{IBM Watson Research Center, PO Box 218, Yorktown Heights, NY 10598, USA}
\affiliation{Institute for Quantum Information, Caltech 107--81,
    Pasadena, CA 91125, USA}
\author{Patrick Hayden}
\email{patrick@cs.caltech.edu}
\affiliation{Institute for Quantum Information, Caltech 107--81,
    Pasadena, CA 91125, USA}
\author{Barbara M. Terhal}
\email{terhal@watson.ibm.com}
\affiliation{Institute for Quantum Information, Caltech 107--81,
    Pasadena, CA 91125, USA}
\affiliation{IBM Watson Research Center, PO Box 218, Yorktown Heights, NY 10598, USA}
\date{\today}

\begin{abstract}
Recent work has shown how to use the laws of quantum mechanics to
keep classical and quantum bits secret in a number of different
circumstances.  Among the examples are private quantum channels,
quantum secret sharing and quantum data hiding.  In this paper we
show that a method for keeping two classical bits hidden in any
such scenario can be used to construct a method for keeping one
quantum bit hidden, and vice--versa.  In the realm of quantum
data hiding, this allows us to construct bipartite and
multipartite hiding schemes for qubits from the previously known
constructions for hiding bits.  
Our method also gives a simple proof
that two bits of shared randomness are required to construct a
private quantum channel hiding one qubit.
\end{abstract}



\maketitle

\section*{Dedication to David Mermin}
\indent {\em  It is a pleasure to have an opportunity to include
our work in this tribute to our friend David Mermin.  We hope
that he will enjoy it, as it is a brand new result flowing from
the teleportation/dense coding mindset that has been so
fantastically productive in quantum information theory over the past
few years. As the now-admitted midwife of teleportation, he will
recognize the usual elements (the ensemble of Pauli rotations, the
one-qubit-gets-you-two-bits-and-vice-versa structure) of these
kinds of arguments, twisted though they may be in the service of
some new cryptographic application. It seems that the metamorphosis of the teleportation
game of 1993 into a myriad of different serious constructions in
the service of cryptography, secure and fault tolerant
computation, and communication complexity has not yet come to an
end.  We are grateful to David Mermin for helping set this process
in motion, and for continuing to observe the resulting flowering
with a generous, interested, and critical eye.}

\section{Introduction}

Work in recent years has shown how to use the laws of quantum
mechanics to keep classical and quantum bits secret in a number of
different circumstances.  In some scenarios, the bits are kept
secret from an eavesdropper while in others, they are kept
secret from the participants themselves.  Perhaps the
simplest such example is the quantum generalization of the
one-time pad, known as a \emph{private quantum channel}
\cite{BR00,AMTW00}.  In this setting, two parties make use of
shared random bits to create a secure quantum channel between
them.  In this case, the message is kept secret from an
eavesdropper with access to the output of the quantum channel.  In
contrast, the goal in \emph{quantum secret sharing} \cite{CGL99}
is to share a secret, in the form of classical or quantum bits,
between many parties.  Certain prescribed combinations of the
parties, known as \emph{authorized sets}, are capable of fully
reconstructing the secret using quantum communication while the
other \emph{unauthorized} combinations of parties can learn
nothing at all about the secret, even if they act jointly on their
shares.  A third example, which will be the focus of this paper,
is known as \emph{quantum data hiding}.  This task, introduced in
Refs. \cite{TDL01,DLT02} for the bipartite setting and generalized
to multiple parties in Ref. \cite{EW02}, imposes a stronger
security criterion than quantum secret sharing.
Whereas in quantum secret sharing an authorized set may be able 
to extract information about the secret by performing local operations in
addition to classical communication, in quantum data hiding
the authorized set needs to communicate quantum data in order to
get substantive information about the secret. So in quantum data hiding one allows all
parties to communicate classical data to one another in an effort
to reveal the secret. Quantum communication within an unauthorized
set, supplemented with classical communication between all
parties, reveals nothing. (Or, rather, next to nothing; one of the
results of Ref. \cite{DLT02} is that perfect quantum data hiding
is impossible.)

\begin{figure}[htb]
\begin{center}
\epsfig{file=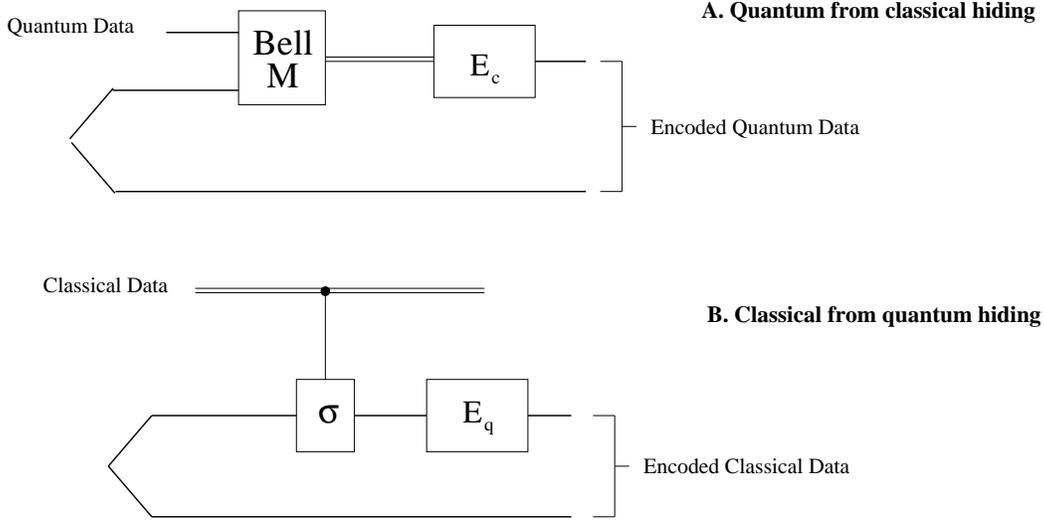,width=14cm} \caption{Conversion between
classical and quantum hiding schemes: A. Using the classical hiding scheme $E_c$ via 
teleportation to get a quantum hiding scheme. B. Using the quantum hiding scheme 
$E_q$ via superdense coding to get a classical hiding scheme.  Double lines denote
classical data and solid lines multi-qubit quantum data. The gate
$\sigma$ applies a (tensor product) of Pauli operations as a
controlled unitary. Bell M denotes a Bell measurement \cite{BBCJPW93}. Note 
that in the 'teleportation' circuit in A, the corrective Pauli
operations are omitted.}
\label{fig:1}
\end{center}
\end{figure}

The main result of this paper is the first construction of quantum
data hiding protocols for hiding qubits; the protocols of Refs.
\cite{DLT02} and \cite{EW02} only work for hiding classical bits. Our
method is to build on top of the earlier work, converting any
method for hiding $2n$ bits into a method for hiding $n$ qubits.
For symmetry, we will also demonstrate how any $n$-qubit data
hiding scheme can be converted into a $2n$-bit hiding scheme.  The
connection, which is closely related to the duality between
superdense coding \cite{BW92} and teleportation \cite{BBCJPW93},
is largely independent of the
setting of the problem.  Indeed, the basic idea, as sketched in
Fig.~\ref{fig:1},  can
actually be applied just as well to the private quantum channel
and quantum secret sharing as to quantum data hiding.

We begin section \ref{sec:bipartite} by defining bipartite data
hiding and describing the method for converting between bit and qubit
hiding schemes.  In sections \ref{sec:QfromC} and \ref{sec:CfromQ} we 
provide security proofs for the resulting schemes under the 
idealized assumption that the original schemes were perfectly secure
before relaxing to approximate hiding in sections \ref{sec:imperfectQ}
and \ref{sec:imperfectC}.
The additional complications that arise in multipartite
hiding are dealt with in section \ref{sec:multi}, where our main result
is that the \emph{only} constraints on the authorized sets are the
same as those for quantum secret sharing \cite{G00}.
Section \ref{sec:resources} demonstrates how the duality imposes limits on the resources
required for hiding.  As an application, we provide a simple,
conceptual proof that $2n$ bits of shared key are required for an
$n$-qubit private quantum channel.


As the reader has probably already noticed, the term \emph{quantum
data hiding} refers to the methods used rather than the data
stored. Rather than resorting to contorted phrases like `quantum
hiding of quantum data', we will henceforth omit the first
`quantum' and refer to \emph{qubit-hiding schemes} or \emph{bit-hiding schemes}. Hopefully, this will simultaneously keep both
confusion and redundancy to a minimum. In this paper we will
denote the density operator corresponding to a pure state
$\ket{\ph}$ as $\ph$. 
The phrase `Trace-preserving Completely Positive map'
will be abbreviated to `TCP map'. A TCP map that can be implemented by
Local Operations supplemented by Classical Communication is called
an LOCC map or operation. The trace norm $||A||_1$ of an operator $A$ is 
defined as $||A||_1={\rm Tr} \sqrt{A^{\dagger}A}$.

\section{Bipartite Hiding}
\label{sec:bipartite}

Formalizing the description of data hiding used in the
introduction, we define an $n$-bit data hiding scheme for two
parties Alice ($A$) and Bob ($B$) to consist simply of a set of
orthogonal bipartite hiding states $\{ \r_I^{AB} \}$, where $I =
i_1 i_2 \dots i_n$ is an $n$-bit string. Since these states are orthogonal, 
we can define a physical encoding map $E_c(\ket{I}\bra{I})=\r_I$. The orthogonality
condition guarantees that, if allowed quantum communication, Alice
and Bob can perfectly recover $I$ by some decoding operations $D$.
The scheme is said to be \emph{perfectly secure} if Alice and Bob
are incapable of learning anything using only LOCC operations. 
Equivalently, the scheme is
perfectly secure if, for all $I$, $J$ and LOCC operations $L$,
\begin{equation}
\Tr_A L(\r_I) = \Tr_A L(\r_J).
\end{equation}
As noted in the introduction, this perfect security is not
actually possible.  We say that the scheme is $\e$-secure if, for
all $I$, $J$ and LOCC operations $L$,
\begin{equation}
\| \Tr_A L(\r_I) - \Tr_A L(\r_J) \|_1 < \e.\label{eqn:e}
\end{equation}
If $\e = 0$ this definition reduces to perfect security. 

Extending this approach to the case of quantum data, we say that a
bipartite $n$-qubit hiding scheme consists of an encoding map
$E_q$ taking $n$-qubit states $\ph$ to bipartite hiding states
$E_q(\ph)$ on $AB$ such that there exists a TCP decoding map $D$
satisfying $D(E_q(\ph)) = \ph$ for all $\ph$.  The scheme is
$\delta$-secure if for all $\ph_0$ and $\ph_1$ as well as LOCC
operations $L$,
\begin{equation}
\| \Tr_A L(E_q(\ph_0)) - \Tr_A L(E_q(\ph_1)) \|_1 < \delta.
\end{equation}
Henceforth we will restrict our attention to pure state inputs
$\ph_0$ and $\ph_1$.  This is sufficient because
the convexity of the trace norm
ensures that the most distinguishable states will always be pure.
For the rest of the paper, we will also impose the additional
requirement that the map $E_q$ correspond to a physical operation,
meaning that it will be TCP. For the qubit-hiding schemes we construct, 
the condition
will be satisfied automatically.  When we attempt to construct bit
hiding schemes from qubit-hiding schemes however, our method would
fail without the extra condition.

Now we are ready to explain how to construct secure hiding
schemes for sets of $n$ qubits starting from secure hiding schemes
for $2n$ classical bits. 

Assume we have a classical hiding scheme for $2n$ bits with encoding
map $E_c$. The hiding states are $\r_I$, where $I =
i_1 i_2 \cdots i_n$ is a string of length $n$, each position
taking an integer value between $0$ and $3$. Let $\ph$ be an
$n$-qubit state. We define a TCP encoding map $E_q$ by
\begin{equation}
E_q(\ph) = \frac{1}{2^{2n}} \sum_I E_c(\ket{I}\bra{I})^{AB_1} \ox
        \s_I \ph \s_I^{B_2}=\frac{1}{2^{2n}} \sum_I \r_I^{AB_1} \ox
        \s_I \ph \s_I^{B_2},\label{eqn:onehide}
\end{equation}
where $\s_I = \s_{i_1} \ox \s_{i_2} \ox \cdots \ox \s_{i_n}$ is a
tensor product of Pauli operators, adopting the convention that
$\s_0 = \id$. It is clear that if $\r_I$ are a set of orthogonal
states, there exists a decoding operation $D$ such that $D \circ
E_q(\ph)=\ph$. Notice that Bob's register is divided into two
parts, the first storing his half of the bit-hiding states $\r_I$
and the second a Pauli-conjugated version of $\ph$. Equivalently,
a more operational way of thinking about $E_q(\ph)$ is as the
output of the circuit illustrated in Fig. \ref{fig:1}A.

Using a dual construction, any perfectly secure $n$-qubit hiding
scheme can be used to build a perfectly secure $2n$-bit hiding
scheme. Let $E_q$ be some TCP encoding map, not necessarily of the
form of Eq. (\ref{eqn:onehide}), hiding states of an $n$ qubit
register $B_1$ on a bipartite system $AB_1$.  (We will assume
without loss of generality when discussing this `superdense coding
scheme' that the initial state is stored on
Bob's system.) Our method of hiding $2n$ classical bits combines the 
quantum hiding scheme and superdense coding, see Fig. \ref{fig:1}B.  Let
\begin{equation}
\ket{\Ph_I}^{B_1 B_2} = (\s_I \ox \id_{B_2}) \ket{\Ph}^{B_1
B_2},\label{eqn:sdc}
\end{equation}
where $\ket{\Ph}^{B_1 B_2} = 2^{-n/2} \sum_{k=1}^{2^n}
\ket{k}^{B_1} \ket{k}^{B_2}$ is a maximally entangled state
between the registers $B_1$ and $B_2$. We define the hiding states
for the classical bits to be
\begin{equation}
\r_I = (E_q \ox \id_{B_2})(\Ph_I^{B_1 B_2}).\label{eqn:hide}
\end{equation}
Since $E_q$ is a qubit-hiding scheme there exists a decoding
operation $D$ such that $D \circ E_q=\id$. This implies that the
classical hiding states $\r_I$ are orthogonal since they can be
mapped by an operation $D \ox \id$ onto the orthogonal states
$\Ph_I$.

In the next few sections we will prove the security of a qubit
hiding scheme from the security of the bit-hiding scheme and
vice-versa.

\subsection{Perfect Hiding: Classical $\rightarrow$ Quantum}
\label{sec:QfromC}

Our goal is to show that if the $\r_I$ are perfectly secure
$2n$-bit hiding states then the states $E_q(\ph)$, defined in Eq. (\ref{eqn:onehide}), 
are, likewise, perfectly secure $n$-qubit hiding states. By construction, some
simple-minded approaches to cheating by Alice and Bob will fail to
yield any information about $\ph$. First, because the reduced
state on the $B_2$ register is always maximally mixed, no
measurement by Bob on $B_2$ alone will yield any information about
the input state. Similarly, since the $AB_1$ register starts
independent of $\ph$, any LOCC protocol applied to it alone will
have output independent of $\ph$. Moreover, since the $\r_I$
form a set of perfect hiding states, Bob's final reduced density
operator on $B_1$ will be independent of $I$ so he can't learn
anything that would help him to undo the $I$-dependent Pauli
rotations on $B_2$.  This doesn't prove security, however.  It is
conceivable that by acting on $B_1$ and $B_2$ together in an LOCC
protocol that Bob might be able cheat by a strategy we haven't yet
considered.  We now give a formal proof that this is not possible.

Suppose, on the contrary, that the proposed $n$-qubit hiding
scheme is \emph{not secure} against arbitrary LOCC cheating.  That
is, there is a choice of input states $\ph_0$ and $\ph_1$ and an
LOCC operation $L$ such that
\begin{equation} \label{eqn:LOCCneq}
\Tr_A L(E_q(\ph_0)) \neq \Tr_A L(E_q(\ph_1)).
\end{equation}
In words, Bob's output density operator at the end of the LOCC
protocol depends on the input state, meaning that he can perform a
local measurement that will distinguish to some degree between
inputs $\ph_0$ and $\ph_1$.   Our goal in what follows will be to
prove that if this were true, the $\r_I$ could not be perfect
hiding states.

For convenience, we'll adopt the more compact notation
$\cL = \Tr_A \circ L$.  We can then introduce the operations
\begin{equation}
\cL_I(\tau^{B_2}) = \cL( \r_I^{AB_1} \ox \tau^{B_2}
)\label{eqn:LI}
\end{equation}
which represent the action of the LOCC operation $\cL$ given
a particular value of the hiding state $\r_I$.  Note that
$\cL_I$, while it is TCP, is \emph{not necessarily} itself an 
LOCC operation because it
involves the preparation of a potentially entangled ancilla
$\r_I^{AB_1}$. We can then write
\begin{equation}
\cL(E_q(\ph)) = \frac{1}{2^{2n}} \sum_I \cL_I(\s_I \ph \s_I)
\end{equation}
for the output of the 'cheating' operation on an $n$-qubit hiding
state.  From this identity, we can conclude that not all the
$\cL_I$ are identical to $\cL_0$, however; if they were,
then by linearity,
\begin{eqnarray}
\cL(E_q(\ph))
&=& \cL_0 \left( \frac{1}{2^{2n}} \sum_I \s_I \ph \s_I \right) \nonumber \\
&=& \cL_0 \left( \frac{1}{2^n} \id_{B_2} \right)
\end{eqnarray}
would be independent of the input state $\ph$, violating
Eq.~(\ref{eqn:LOCCneq}).

The non-constancy of the $\cL_I$ can then be converted into a
method for breaking the $2n$-bit hiding scheme.  Supplied with
a state $\r_I^{AB_1}$ from which they would like to learn about
$I$, Alice and Bob implement the following LOCC protocol.  First,
Bob prepares a maximally entangled state
\begin{equation}
\ket{\Ph}^{B_2 B_3}
= \frac{1}{\sqrt{2^n}} \sum_{k=1}^{2^n} \ket{k}^{B_2} \ket{k}^{B_3}
\end{equation}
between two local registers $B_2$ and $B_3$. Alice and Bob then
together apply the LOCC operation $\cL \ox \id_{B_3}$ to the state
$\r_I^{AB_1} \ox \Ph^{B_2 B_3}$, resulting in the outcome $(\cL_I
\ox \id_{B_3})( \Ph )$ on Bob's system alone.  By the
Jamio{\l}kowski isomorphism between operations and states
\cite{J72}, the non-constancy of the $\cL_I$ implies that the
outcome cannot be independent of $I$.  Hence, Bob can perform a
local measurement whose outcome will be $I$-dependent and the $2n$
bit hiding scheme based on the $\r_I$ cannot be secure.

\subsection{Perfect Hiding: Quantum $\rightarrow$ Classical} \label{sec:CfromQ}

Consider the definition of the bit-hiding states in Eq. (\ref{eqn:hide}). 
Because we can choose an operator
basis consisting of density operators $\tau_J$, there is an
expansion
\begin{equation} \label{eqn:projExpansion}
\proj{\Ph}^{B_1B_2} = \sum_{JK} \a_{JK} \tau_J \ox \tau_K
\end{equation}
of the projector for the maximally entangled state in terms
of density operators for product states.  (The $\a_{JK}$, of course,
will not all be positive.)

Now let's try cheating on our $2n$-bit hiding scheme. If $\cL =
\Tr_A \circ L$ is again an LOCC operation with output on Bob's
system, then substituting Eq.~(\ref{eqn:projExpansion}) into
(\ref{eqn:hide}) shows that
\begin{eqnarray}
\cL( \r_I )
&=& \cL( (E_q \ox \id_{B_2})( \Ph_I^{B_1B_2} ) ) \nonumber \\
&=& \sum_{JK} \alpha_{JK} \cL( E_q(\s_I \tau_J \s_I)^{AB_1} \ox
\tau_K^{B_2} ).
\end{eqnarray}
The operation of first preparing $\tau_K$ on $B_2$ and then
applying $\cL$ is itself LOCC so by the perfect security of the
$n$-qubit hiding scheme, we can conclude that $\cL(E_q(\s_I \tau_J
\s_I)^{AB_1} \ox \tau_K^{B_2})$ is independent of $\s_I \tau_J
\s_I$ for all $I$ and $J$.  Consequently, $\cL( \r_I )$ is
independent of $I$, meaning that the $2n$-bit hiding scheme is
perfectly secure.

\subsection{Imperfect Hiding: Classical $\rightarrow$ Quantum}
\label{sec:imperfectQ}

As was shown in Ref. \cite{DLT02}, while a
bit-hiding scheme can be made $\e$-secure for all $\e>0$, perfect
security is not possible. So, we need to investigate whether a
nearly secure bit-hiding scheme leads to a nearly secure qubit-hiding 
scheme.  We show that an $\e$-secure $2n$-bit hiding scheme
can be converted into a $\d$-secure $n$-qubit hiding scheme, for
$\d = \e 2^{n+1}$. The exponential factor $2^{n+1}$, while
undesirable, needn't cause practical difficulties: for the
bit-hiding schemes presented in Refs. \cite{DLT02} and
\cite{EW02}, $\epsilon$ decreases exponentially with the size of
the hiding state. Therefore, the factor $2^{n+1}$ can be
suppressed at a cost of increasing the size of the hiding state by
a factor polynomial in $n$. Furthermore, it could very well be
that the estimates we present here are not tight and that the
factor is only an artifact of our analysis. In any case, the idea
behind the proof of security is the same as in the perfect case
but the details, unfortunately, become significantly more
technical. So as not to repeat ourselves, we will adopt the
notation of section \ref{sec:QfromC}.

In the perfect case, we proceeded by making a connection between
the behavior of $\cL \circ E_q$ and $(\cL_I \ox
\id_{B_3})(\Ph^{B_2 B_3})$. Name this last operator $\o_I$ and
introduce $\D_I = \o_I - \o_0$. Recall also that the encoding
operation $E_q$ takes $B_2$, an $n$-qubit system, to $AB_1 B_2$
while the LOCC $\cL$ takes $AB_1 B_2$ to a Bob-only system. Since
we don't want to make any assumptions yet about it's structure, we
will call this system $B_f$. We define the state
\begin{equation}
\xi^{B_f B_3}=\left((\cL \circ E_q) \ox \id_{B_3}\right)(\Ph^{B_2 B_3}).
\end{equation}
The state $\xi$ can be related to the
action of the map $\cL \circ E_q$ on a state $\ph$ by the
identity
\begin{equation}
\ph=2^n {\rm Tr}_2 ((\id \otimes \ph*) \Ph^{12}),
\label{eqid}
\end{equation}
where $\ph^*$ is the complex conjugate of the density matrix of $\ph$
and the numbers $1$ and $2$ are general system labels. 
Now, as in the perfect hiding case, assume that the
$n$-qubit hiding scheme is not $\d$-secure, meaning that there exist
states $\ph_0$ and $\ph_1$ such that
\begin{equation}
\d < \| ( \cL \circ E_q )(\ph_0) - (\cL \circ E_q)( \ph_1 )
\|_1.\label{eqn:contra}
\end{equation}
Using the identity in Eq. (\ref{eqid}) and the definition of
$\xi$, we find that
\begin{eqnarray}
\d &<& 2^n \| \Tr_{B_3} (\id \ox \ph_0^*) \xi^{B_f B_3}
    - \Tr_{B_3} (\id \ox \ph_1^*) \xi^{B_f B_3} \|_1.
\label{delta}
\end{eqnarray}
In order to relate $\d$ to $\Delta_I$, we rewrite $\xi$ in the following manner: 
\begin{eqnarray} \label{eqn:xiDecomp}
\xi^{B_f B_3}
&=& \left( (\cL \circ E_q) \ox \id_{B_3} \right) \Ph^{B_2 B_3} \nonumber \\
&=& \frac{1}{2^{2n}} \sum_I ({\cL}_I \ox \id_{B_3})
    \left( (\s_I \ox \id_{B_3}) \Ph^{B_2 B_3} (\s_I \ox \id_{B_3}) \right) \nonumber \\
&=& \frac{1}{2^{2n}} \sum_I (\id_{B_f} \ox \s_I) \o_I (\id_{B_f} \ox \s_I) \nonumber \\
&=& \Tr_{B_3} \o_0 \ox \frac{1}{2^n} \id_{B_3} +
    \frac{1}{2^{2n}} \sum_I (\id_{B_f} \ox \s_I) \D_I (\id_{B_f} \ox \s_I),
\end{eqnarray}
where we have used the fact that $(\id \ox \s_I)\ket{\Ph} = \pm
(\s_I \ox \id)\ket{\Ph}$. When inserting this in Eq. (\ref{delta}) we observe 
that the term involving $\Tr_{B_3} \o_0 \ox \frac{1}{2^n} \id_{B_3}$ makes no
contribution so we need only keep the sum over
$(\id\ox\s_I)\D_I(\id\ox\s_I)$. In the following derivation we will need the inequality
\begin{equation}
||P A||_1 \leq ||A||_1,
\label{pro}
\end{equation}
where $P$ is a projector. This can be proved as follows. 
Let $\lambda_1 \geq \lambda_2 \geq \ldots$ be the singular 
values of $A$. Let $P$ be a $k$-dimensional projector. We have 
\begin{equation}
||PA ||_1=\max_U {\rm Tr} (PAU) \leq 
\max_{U,Q} {\rm Tr} (QAU)=\sum_{i=1}^k \lambda_i \leq ||A||_1,
\end{equation}
where we used that $Q$ is $k$-dimensional projector.

We insert the result of Eq. (\ref{eqn:xiDecomp}) in Eq. (\ref{delta}) and 
apply Eq. (\ref{pro}), the monotonicity under partial trace and subadditivity of the trace norm to find
\begin{eqnarray}
\d
&<& \frac{1}{2^n} \sum_I \left\|
    (\id \ox \ph_0^* )
    (\id \ox \s_I) \D_I (\id \ox \s_I)
    \right\|_1  \nonumber \\
&\;&    +\frac{1}{2^n} \sum_I \left\|
    (\id \ox \ph_1^*)
    (\id \ox \s_I) \D_I (\id \ox \s_I)
    \right\|_1 \nonumber \\
&\leq& \frac{1}{2^{n-1}} \sum_I \left\|
    (\id \ox \s_I) \D_I (\id \ox \s_I) \right\|_1 \nonumber \\
&=& \frac{1}{2^{n-1}} \sum_I \left\| \D_I
\right\|_1.\label{eqn:subad}
\end{eqnarray}
Reading this inequality as an average over the $2^{2n}$ possible values
of $I$, there must exist a particular choice for $I$ for which
$\|\o_{I} - \o_{I=0} \|_1 = \|\D_{I}\|_1 > \d/2^{n+1}$.  As
in the perfect hiding argument, this provides
a cheating operation for the classical scheme that will distinguish
the hiding states of the particular $I$ and $I=0$.  The classical scheme,
therefore, cannot be $\d/2^{n+1}$-secure.

\subsection{Imperfect Hiding: Quantum $\rightarrow$ Classical}
\label{sec:imperfectC}

We suppose that there is some quantum hiding scheme $E_q$ that 
is $\delta$-secure, i.e. for all pairs of quantum states 
$\ph_0$ and $\ph_1$ and LOCC operations $\cL$ we have 
\begin{equation}
\|(\cL \circ E_q)(\ph_0)-(\cL\circ E_q)(\ph_1)\|_1\leq\delta.
\label{bdelta}
\end{equation}
>From this we will deduce the quality of the derived bit-hiding
scheme, that is, we study
\begin{equation}
\|{\cal L}(\rho_I)-{\cal L}(\rho_J)\|_1,\label{eqn:target}
\end{equation}
where $\rho_I$ and $\rho_J$ are given in Eq.
(\ref{eqn:hide}). We will use an explicit operator expansion of
the maximally entangled projector Eq. (\ref{eqn:projExpansion})
\cite[Eq.~(24)]{DLT02}:
\begin{equation}
\Ph=\frac{1}{4^n}
\sum_{M=0}^{4^n-1}(-1)^{N(11)}\sigma_M\otimes\sigma_M.
\end{equation}
Here $N(11)$ counts the number of $\sigma_y$ operators in the
product $\sigma_M$.  Note also that $\sigma_M$, for any $M$, has
$2^{n-1}$ positive eigenvalues ($\lambda=+1$), and the same number
of negative eigenvalues (
$\lambda=-1$); therefore, it can be written as the
difference of two density operators using
\begin{equation}
\sigma_M=2^{n-1}(\rho_+^M-\rho_-^M),
\end{equation}
where $\rho_{\pm}^M$ are separable. In the following, we will use the shorthand
$\sigma_{IMI}=\sigma_I\sigma_M\sigma_I$, and
$\rho_\pm^{IMI}=\sigma_I\rho_\pm^M\sigma_I$.  With all this, we
can write for Eq. (\ref{eqn:target}):
\begin{eqnarray}
\|{\cal L}(\rho_I)-{\cal L}(\rho_J)\|_1&&=\left\|{\cal L}
((E_q\otimes \id_{B_2})(\proj{\Ph_I}-\proj{\Ph_J}))\right\|_1\nonumber\\
&&=\left\|\frac{1}{4^n}\sum_M(-1)^{N(11)}[{\cal
L}(E_q(\sigma_{IMI})\otimes\sigma_M)-{\cal
L}(E_q(\sigma_{JMJ})\otimes\sigma_M)]\right\|_1\nonumber\\
&&\leq\frac{1}{4^n}\sum_M\left \| {\cal
L}(E_q(\sigma_{IMI})\otimes\sigma_M)\right \|_1+\left \| {\cal
L}(E_q(\sigma_{JMJ})\otimes\sigma_M)\right\|_1\nonumber\\
&&\leq\frac{2}{4^n}\max_K \sum_M\left \| {\cal
L}(E_q(\sigma_{KMK})\otimes\sigma_M)\right \|_1\nonumber\\
&&\leq\frac{2\cdot 2^{2(n-1)}}{4^n}\max_K \sum_M\left\| {\cal
L}(E_q(\rho_+^{KMK})\otimes\rho_+^M) -{\cal
L}(E_q(\rho_-^{KMK})\otimes\rho_+^M) +\right .\nonumber\\
&&\left . \,\,\,\,{\cal L}(E_q(\rho_-^{KMK})\otimes\rho_-^M)
-{\cal L}(E_q(\rho_+^{KMK})\otimes\rho_-^M)\right \|_1
\nonumber\\
&&\leq\frac{1}{2}\max_K \sum_M\left\| {\cal
L}(E_q(\rho_+^{KMK})\otimes\rho_+^M) -{\cal
L}(E_q(\rho_-^{KMK})\otimes\rho_+^M)\right\|_1+\nonumber\\
&&\,\,\,\,\frac{1}{2}\max_K \sum_M\left \|{\cal
L}(E_q(\rho_-^{KMK})\otimes\rho_-^M) -{\cal
L}(E_q(\rho_+^{KMK})\otimes\rho_-^M)\right \|_1
\nonumber\\
&&\leq\frac{4^n}{2}\max_K \max_M\left\| {\cal
L}(E_q(\rho_+^{KMK})\otimes\rho_+^M) -{\cal
L}(E_q(\rho_-^{KMK})\otimes\rho_+^M)\right\|_1+ \nonumber
\\
&&\,\,\,\,\frac{4^n}{2}\max_K \max_M\left \|{\cal
L}(E_q(\rho_-^{KMK})\otimes\rho_-^M) -{\cal
L}(E_q(\rho_+^{KMK})\otimes\rho_-^M)\right \|_1
\nonumber\\
&&\leq\frac{4^n}{2}(2\delta)=4^n\,\delta.
\end{eqnarray}
In the last inequality, we used the fact that $\rho_{\pm}^M$ are separable density
matrices, independent of $K$, which implies that Alice and Bob can prepare them by LOCC operations.
Thus to distinguish, say, $E_q(\rho_+^{KMK}) \otimes \rho_+^M$ from $E_q(\rho_-^{KMK}) \otimes \rho_+^M$
 by LOCC should not be easier then to distinguish $E_q(\rho_+^{KMK})$ from $E_q(\rho_-^{KMK})$ 
 by LOCC, for which the distinguishability is bounded as in Eq. (\ref{bdelta}). 

So, we get a bound on the quality of the bit-hiding scheme,
although one suffering the same exponential deficiency as the bound
of section \ref{sec:imperfectQ}.

\section{Multiparty hiding \& quantum secret sharing} \label{sec:multi}

We now consider the task of hiding quantum data in a multiparty setting.
Generalizations of bipartite bit-hiding schemes to multipartite 
situations have been developed by Eggeling and Werner \cite{EW02}.
Unfortunately, there is a problem with combining these bit-hiding schemes
with the qubit-hiding construction of Eq.~(\ref{eqn:onehide}): the
qubit-hiding scheme places the hidden quantum state entirely in the
possession of a single party since register $B_2$ belongs to Bob.  
In the direct generalization of the scheme to the multiparty setting,
the privileged holder of the hidden quantum state would, therefore, 
necessarily have to be a member of every authorized set.  This would
eliminate the possibility of threshold schemes, for example, in which
any sufficiently large subset of the parties should be able reconstruct the 
secret.  The solution is to hide \emph{distributed} quantum data, using
quantum error correcting codes to share the hidden quantum state between
the parties in a more symmetrical fashion.

An application involving such distributed hiding has been considered in
the literature: quantum secret sharing \cite{CGL99}.  In quantum
secret sharing, the identity of a distributed quantum state, held
by a set of parties, is unobtainable by these parties if they do
only local operations. There are {\em authorized sets} of parties who, with quantum
communication among each other, can reconstruct (i.e., put the
full state in the possession of any single party) the quantum
state, and there are unauthorized sets, for whom no reconstruction
is possible even with quantum communication. Such quantum secret
sharing schemes are implementable with quantum error correcting
codes; for example, there is a five-qubit error correcting code
for which any three out of the five parties constitute an
authorized set, while any set of two is unauthorized.  With such
error correction codes, quantum secret sharing schemes with any
``access structure" are realizable.  This access structure need
only be consistent with monotonicity and the quantum no-cloning
theorem, meaning that any superset of an authorized set is
authorized, and the complement of an authorized set is
unauthorized \cite{G00}.

\begin{figure}[htb]
\begin{center}
\epsfig{file=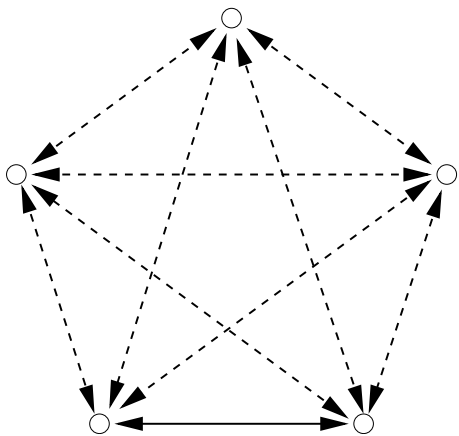,width=6cm} \caption{If five parties each
have shares of a state in a five-qubit error correcting code, then
the identity of the state cannot be obtained by two parties in
quantum communication \cite{G00}.  But if all parties are in
addition in classical communication with one another (dotted
lines), then the parties can obtain some information about the
state.  If, however, the secret-sharing state is encoded using a
bit-hiding state holding $2\times 5$ bits (Eq. (\ref{multihide})),
then the parties can obtain only a negligible amount of
information about the hidden quantum state.} \label{fig:2}
\end{center}
\end{figure}

The capabilities of quantum secret sharing can be strengthened by
the techniques of this paper.  The quantum secret sharing protocol
does not specify the status of the secret if the parties can
perform LOCC operations, rather than just local operations.  In
fact, for the implementation of quantum secret sharing using
quantum error correcting codes, LOCC operations between the
parties can, and often do, result in the parties obtaining partial
information about the secret. 
However, by wrapping the quantum
state of quantum secret sharing inside a multipartite version of
our qubit-hiding protocol, we can guarantee that the
quantum secret is impervious to attack by LOCC operations of the
parties; we illustrate the idea in Fig. \ref{fig:2}.  This
requires some generalization of the protocol given above, and of
its security analysis.

Suppose we have a $p$-party quantum secret sharing state $\ph$,
and that $k$ qubits distributed among these parties are sufficient
to hold this state.  (For quantum secret sharing schemes, $k$ is
polynomially related to the number of logical qubits $n$ that can
be hidden in such a state, $k=poly(n,p)$.) Now, we can create a
new state with stronger security properties using the map
\begin{equation}
E(\ph)=\frac{1}{2^{2k}}
\sum_I\rho_I^1\otimes(\sigma_I\ph\sigma_I)^2\label{multihide}
\end{equation}
Here the quantum secret lives in subsystem ``2", which is an
$k$-qubit multipartite Hilbert space distributed among the $p$
parties ${\cal H}_{A_2}\otimes{\cal H}_{B_2}\otimes{\cal
H}_{C_2}\otimes ...$. The dimensions of these local spaces need not be 
the same, since some parties may get larger shares of the secret than others.
Each party also has a register of subsystem ``1", comprising $s$ qubits in total, 
which contains the data hiding state $\rho_I$, and which is capable of hiding the
$2k$-bit string $I$.  Note that, although the secret-sharing state
$\ph$ may only occupy a subspace of the ``2" subsystem (as when it
is a quantum error correcting code state), $\sigma_I$ acts on the
entire ``2" Hilbert space, not just on the code subspace in which
the quantum secret may be contained.

As Eggeling and Werner have recently shown, there exist
multipartite data-hiding states $\rho_I$ with any desired access
structure, and with hiding security that is exponential in $s/k$ \cite{EW02}.
(Unlike the bipartite states used in Ref.~\cite{DLT02}, however, these
states are not orthogonal, just nearly so.  This gives rise to a small
probability of error when authorized sets reconstruct the secret
but otherwise has no effect on the analysis for our purposes.)
If we choose the access structure of the $\rho_I$ states and the
$\ph$ quantum secret sharing states to be identical, then the
quantum state can obviously still be reconstructed by quantum
communication within an authorized set; first $\rho_I$ is measured
to identify $I$, then the Pauli rotation $\sigma_I$ is done by the
authorized parties, so that they have the state $\ph$ ``in the
clear", permitting them to reconstruct it by whatever operations
the original quantum secret sharing protocol prescribed. Of
course, it does not matter whether the Pauli rotations are done by
the parties outside the authorized set, since these parts of the
quantum state $\ph$ are not needed for the reconstruction anyway.

In the rest of this section, we demonstrate the other part of the
desired security of the protocol: an unauthorized set cannot reveal 
the quantum state even when all parties can perform LOCC and 
quantum communication can be performed within the unauthorized set.
The proof relies on the fact, as in the bipartite case, that with these 
resources, the parties cannot decode the classical hiding states $\r_I$.

So, we suppose that in the given ``unauthorized" setting, the
$2k$-bit hiding scheme is $\epsilon$-secure; we will show that the
quantum-state hiding is guaranteed to be $\delta$-secure, for
$\delta=\epsilon 2^{3k+5}$.  The first part of the demonstration
closely follows the reasoning of Section \ref{sec:imperfectQ}.  We
consider ``cheating" operations ${\cal L}^+={\rm Tr}'
\circ L^+$, where $L^+$ is a member of the set of LOCC operations + quantum
operations among members of the unauthorized set, and ${\rm Tr}'$
indicates a tracing out of all parties except one.  We will also
need the multipartite version of Eq. (\ref{eqn:LI}), ${\cal
L}_I^+(\tau^2)={\cal L}^+(\rho_I^1\otimes\tau^2)$.  Given that the
parties are supplied with the state $\rho_I^1$, ${\cal L}_I^+$ can
be implemented with the same limited communication resources as
${\cal L}^+$ can.  We introduce ancilla subsystem ``3", which has
the same dimension and the same multipartite structure as
subsystem ``2". Each party locally creates a maximally entangled state 
$\Phi$ between its part of system 2 and 3. Let us denote the tensor product 
of all these local maximally entangled states as a big maximally entangled 
state $\Phi^{23}$. Then the parties can create the state
\begin{equation} \label{eqn:multiOm}
\omega_I=({\cal L}_I^+ \otimes \id_3)(\Phi^{23}).
\end{equation}

Now the proof begins by contradiction as in Section
\ref{sec:imperfectQ}: suppose the qubit-hiding scheme is not
$\delta$-secure, meaning that there are secret sharing states
$\ph_0$ and $\ph_1$ and a cheating operation ${\cal L}^+$ such
that Eq. (\ref{eqn:contra}) is true.  Then by following without
change the analysis after Eq. (\ref{eqn:contra}), we conclude that
there must exist a bit string $I$ for which (cf. below
Eq. (\ref{eqn:subad}))
\begin{equation}
\|\o_{I=0} - \o_{I} \|_1 = \|\D_{I}\|_1 >
\d/2^{k+1}.\label{eqn:oldb}
\end{equation}
Unlike in the previous case, we cannot use this equation
immediately to bound $\epsilon$ and end the argument; the trace
norm is only directly related to the distinguishability when any
quantum operation can be done on the state.  In this case
$\omega_I$ is a state shared by all parties who can perform LOCC
operations and some additional quantum communication depending on
the protocol. In other words, we can only bound the
distinguishability $\epsilon$ of the classical message by (see Eq.
(\ref{eqn:e}))
\begin{equation}
\epsilon\geq\max_{I,\,K}{\rm Dist}_{{\cal
L}^+}(\omega_{I},\omega_{K})\geq\max_{I,\,K}{\rm Dist}_{{\cal
L}_{LOCC}}(\omega_{I},\omega_{K})=\max_{I,\,K}\max_{{\cal
L}_{LOCC}}\|{\cal L}_{LOCC}(\omega_{I})-{\cal
L}_{LOCC}(\omega_{K})\|_1.\label{thing0}
\end{equation}
Here ${\rm Dist}_X$ denotes the distinguishability of two states
under the set of operations $X$.  We restrict to only LOCC
operations here because we can use a known relationship between
the LOCC distinguishability of two states and their trace-norm
distance, using the tomography arguments of Ref. \cite{DLT02}.  Of
course, this LOCC distinguishability may be very much less than
$\|\omega_{I=0}-\omega_{I}\|_1$, precisely because of the
data-hiding effect that is the subject of this paper, which
sometimes prevents the distinguishability of states from being
detectable by LOCC operations. The effect is never perfect,
however, a fact which we now use.

Suppose the states $\omega_I$ are written
as~\cite[Eq.~(105)]{DLT02}
\begin{equation}
\omega_{I}=\frac{1}{d}\sum_{J}a_{IJ}\sigma_{J},
\end{equation}
where $d$ is the dimension of the space supporting $\o_I$.
The ``3'' register in Eq.~(\ref{eqn:multiOm}) supports $k$ qubits
whereas the map $\cL_I$ can be taken to output a single 
qubit \footnote{The quantum operation with single-bit outcome corresponding 
to projection onto the positive and negative subspaces of $\tau_0 - \tau_1$
for any density operators $\tau_i$ will have trace norm distance between
the outcomes exactly equal to $\|\tau_0 - \tau_1\|_1$.}.  
Therefore, $d=2^{k+1}$.  Then Appendix B of Ref. \cite[Eq.~(110)]{DLT02} shows that
\begin{equation}
{\rm Dist}_{LOCC}(\omega_{I},\omega_{K})\geq\frac{1}{2}
\max_{J}|a_{IJ}-a_{KJ}|. \label{thing1}
\end{equation}
We apply a chain of inequalities:
\begin{eqnarray}
&&\|\omega_{I}-\omega_{K}\|_1\leq\frac{1}{d}\sum_{J}
\|(a_{IJ}-a_{KJ})\sigma_{J}\|_1\leq\frac{1}{d}\sum_{J}
\|a_{IJ}-a_{KJ}\|_1\|\sigma_{J}\|_1\nonumber\\
&&=\sum_{J}|a_{IJ}-a_{KJ}|\leq
d^2\max_{J}|a_{IJ}-a_{KJ}|.\label{thing2}
\end{eqnarray}
We have used $||\sigma_{J}||_1=d$.  Note also that there are $d^2$
terms in the $J$-sum.  Combining (\ref{thing0}), (\ref{thing1}) and
(\ref{thing2}):
\begin{equation}
\epsilon\geq\frac{1}{2 d^2}||\omega_{I=0}-\omega_I||_1.
\end{equation}
So, combining this with Eq. (\ref{eqn:oldb}), we find that
\begin{equation}
\epsilon\geq\frac{\delta}{4d^3}=\frac{\delta}{2^{3k+5}}.
\end{equation}
But since the classical hiding can be chosen such that
$\epsilon=c_12^{-c_2 s/k}$, there is always a choice of $s$ that
will guarantee that $\delta$ is as small as desired.  So, at the
price of a worse bound (but only polynomially worse), we prove
security of the multipartite case.

\section{How many classical bits are needed in quantum hiding} \label{sec:resources}

In the previous sections we have seen that any $2n$-bit hiding
scheme can be used to construct an $n$-qubit hiding scheme and
vice-versa.  This duality has immediate implications for the
resource requirements of quantum data hiding schemes.  In
particular, suppose that $\{\r_I\}$ is a set of perfectly secure
hiding states for the string $I$, representing $k$ bits of data
and that
\begin{equation} \label{eq:genQfromC}
E_q(\ph) = \frac{1}{2^k} \sum_I \r_I^{AB_1} \ox T_I(\ph)^{B_2},
\end{equation}
where $T_I$ is a TCP map and $\ph$ is an $n$-qubit state. We do
not know whether all $n$-qubit hiding schemes will have this form
but it is a significant generalization of the construction we
described in section \ref{sec:bipartite}. We will show that in order 
for this to be a secure qubit hiding scheme $k \geq 2n$.

Let us assume that this provides a perfectly secure $n$-qubit 
hiding scheme for $\ph$, and use the encoding $E_q$ to hide bits by 
means of our superdense coding construction.
We will get a secure $2n$-bit hiding scheme by applying $E_q\ox \id$ 
to the appropriate maximally entangled states. 
We could interpret this construction as a way of hiding a message
of $2n$ bits by means of a key $I$ of $k$ bits. We will now prove that
this implies that $k \geq 2n$, using an argument nearly identical to
the one Shannon used to show that one-time pad encryption of a $2n$-bit
message requires $2n$ shared random key bits \cite{S49}.  
The only difference
here is that we substitute quantum entropy functions for their classical
counterparts and then have to verify in a couple of places that these quantum
functions are nonnegative, a property guaranteed for their classical 
versions.  For definitions of the functions we use below, see,
for example, Ref.~\cite{NC00}.

Consider the density operator
\begin{equation}
\sum_{m,I} p_m \proj{m}^M \ox \proj{I}^K \ox \r_I^{AB_1}
    \ox (T_I \ox \id)(\proj{\Ph_m})^{B_2 B_3}.
\end{equation}
Here $M$ is a register storing the message $m$, $K$ a register
storing the key $I$ and the set $\{\Ph_m\}$ is a set of mutually
orthogonal maximally entangled states.  Because the message and
key are independent, $S(M:K)=0$.  Likewise, because the bit-hiding
scheme is perfectly secure, $S(M:B_2 B_3)=0$.  Once the key is known,
however, the classical message can be reconstructed from register
$B_2 B_3$ so that
\begin{equation}
S(M:B_2 B_3|K)=S(M:K|B_2 B_3)=S(M).
\end{equation}
Equivalently, $S(K|B_2 B_3)-S(K|B_2 B_3 M ) = S(M)$. Because the
multipartite density operator is separable across the $M/K/AB_1/B_2 B_3$
cuts, $S(K|B_2 B_3 M) \geq 0$, and we can conclude
that $S(K)\geq S(M)$.  In particular, applying this inequality for
the uniform distribution over $2n$ bit messages yields $k \geq
2n$.  Our conclusion is that if an $n$-qubit hiding scheme is constructed from a $k$-bit
hiding scheme in the manner of Eq.~(\ref{eq:genQfromC}) then $k$,
the number of bits, must be at least twice the number of qubits being
hidden.

These arguments can also be applied to the case of a private quantum channel. 
In this scenario, the analog of an $k$-bit
hiding scheme is just $k$ secret random bits shared between two
parties Alice and Bob.  A general private quantum channel then has
the form
\begin{equation}
E(\ph) = \frac{1}{2^k} \sum_I \proj{I,I}^{AB} \ox T_I(\ph)^C,
\end{equation}
where $T_I$ is a general TCP map with output on the channel
system $C$.  The requirements for the task are that, using their
access to $I$, Bob (or Alice) can reconstruct $\ph$ but an
eavesdropper with access only to the channel $C$ can learn nothing.
Assume that one can encrypt an $n$-qubit quantum state $\ph$ in this manner.
Such a private quantum channel can be converted into a secure
$2n$-bit one-time pad using the superdense coding construction of
section \ref{sec:CfromQ} -- the proof goes through unchanged.  
Likewise, the resource considerations developed above imply that $k \geq 2n$
bits of shared secret key are necessary and sufficient to build
the $n$-qubit private quantum channel, confirming the main result
of Refs. \cite{BR00} and \cite{AMTW00}.


\section{Discussion and conclusions}

Our main goal in this paper was to show how the duality between
superdense coding and teleportation can be used to construct new
cryptographic protocols.  From a constructive point of view, our
main results are a protocol for hiding qubits given a protocol for
hiding twice as many bits and a method for strengthening quantum
secret sharing protocols such that they are not vulnerable to
cheating by LOCC.

In our analyses of imperfect hiding, however, we could only guarantee
a quality of hiding decreasing exponentially in the number of
qubits, or bits, being hidden.  Because the quality typically
\emph{improves} exponentially with the size of the hiding state
measured in qubits, our security proofs could still be useful.
Nonetheless, it seems possible that a more careful analysis of the
hiding quality of the new protocols would reveal that the
exponential factor, in fact, disappears.  We leave that possibility
open for future work.

\subsection*{Acknowledgments}

We thank Dave Bacon, Debbie Leung and Andreas Winter for discussions of 
this work.
DDV is grateful for the support of the National Security Agency
and the Advanced Research and Development Activity through Army
Research Office contract number DAAD19-01-C-0056, and for the
support of the National Reconnaissance Office.  BMT and PH
acknowledge support from the National Science Foundation under
Grant. No. EIA-0086038.  PH's work is also supported by a Sherman
Fairchild Fellowship.




\bibliographystyle{unsrt}
\bibliography{qqhiding}

\begin{thebibliography}{10}

\bibitem{BR00}
P.~O. Boykin and V.~Roychowdhury.
\newblock Optimal encryption of quantum bits.
\newblock {LANL} e-print quant-ph/0003059.

\bibitem{AMTW00}
A.~Ambainis, M.~Mosca, A.~Tapp, and R.~de~Wolf.
\newblock Private quantum channels.
\newblock In {\em {IEEE} Symposium on Foundations of Computer Science (FOCS)},
  pages 547--553, 2000.
\newblock {LANL} e-print quant-ph/0003101.

\bibitem{CGL99}
R.~Cleve, D.~Gottesman, and H.~K. Lo.
\newblock How to share a quantum secret.
\newblock {\em Physical Review Letters}, 83(3):648--651, 1999.
\newblock {LANL} e-print quant-ph/9901025.

\bibitem{TDL01}
B.~M. Terhal, D.P. DiVincenzo, and D.~W. Leung.
\newblock Hiding bits in {B}ell states.
\newblock {\em Physical Review Letters}, 86(25):5807--5810, 2001.
\newblock {LANL} e-print quant-ph/0011042.

\bibitem{DLT02}
D.P. DiVincenzo, D.~W. Leung, and B.~M. Terhal.
\newblock Quantum data hiding.
\newblock {\em {IEEE} Transactions on Information Theory}, 48(3):580--598,
  2002.
\newblock {LANL} e-print quant-ph/0103098.

\bibitem{EW02}
T.~Eggeling and R.~F. Werner.
\newblock Hiding classical data in multi-partite quantum states.
\newblock {LANL} e-print quant-ph/0203004.

\bibitem{BW92}
C.~H. Bennett and S.~Wiesner.
\newblock Communication via one- and two-particle operators on
  {E}instein-{P}odolsky-{R}osen states.
\newblock {\em Physical Review Letters}, 69(20):2881--2884, 1992.

\bibitem{BBCJPW93}
C.~H. Bennett, G.~Brassard, C.~Cr\'{e}peau, R.~Jozsa, A.~Peres, and W.~K.
  Wootters.
\newblock Teleporting an unknown quantum state via dual classical and
  {E}instein-{P}odolsky-{R}osen channels.
\newblock {\em Physical Review Letters}, 70:1895--1899, 1993.

\bibitem{G00}
D.~Gottesman.
\newblock Theory of quantum secret sharing.
\newblock {\em Physical Review A}, 61(4):042311, 2000.
\newblock {LANL} e-print quant-ph/9910067.

\bibitem{J72}
J.~Jamio{\l}kowski.
\newblock Linear transformations which preserve trace and positive
  semidefiniteness of operators.
\newblock {\em Reports on Mathematical Physics}, 3(4):275--278, 1972.

\bibitem{S49}
C.~E. Shannon.
\newblock Communication theory of secrecy systems.
\newblock {\em Bell System Technical Journal}, 28:656--715, 1949.

\bibitem{NC00}
M.~A. Nielsen and I.~L. Chuang.
\newblock {\em Quantum computation and quantum information}.
\newblock Cambridge University Press, 2000.

\end{thebibliography}

\end{document}